\documentclass[11pt]{article}
\usepackage{color}
\usepackage{amsmath,amssymb}
\usepackage{graphicx} 
\usepackage{bbm}
\usepackage{longtable}
\usepackage[small,loose]{subfigure}
\usepackage[small]{caption2} 
\usepackage{fleqn} 
\usepackage{cite}
\usepackage{pifont}

\advance \headheight by 3.0truept       
\setcaptionwidth{0.85\textwidth}

\allowdisplaybreaks

\begin{document}

\date{}
\title{
\begin{flushright}
\normalsize{DESY 10-198}\\
\end{flushright}
\vskip 2cm
{\bf\huge CP violation with Higgs--dependent Yukawa couplings}\\[0.8cm]}

\author{{\bf\normalsize
Oleg~Lebedev }\\[1cm]
{\it\normalsize
DESY Theory Group}\\ 
{\it\normalsize Notkestrasse 85, D-22603 Hamburg, Germany
}
}

\maketitle 

\thispagestyle{empty}

\vskip 1cm
\begin{abstract}
 The framework of Higgs--dependent Yukawa couplings allows one to eliminate 
small couplings from the Standard Model, which can be tested at the LHC. 
In this work, I study the conditions for CP violation to occur in such models.
I identify a class of weak basis invariants controlling CP violation.
The invariant measure of CP violation is found to be more than 10 orders 
of magnitude greater than that in the Standard Model, which can be sufficient 
for successful electroweak baryogenesis.  
\end{abstract}
\clearpage

\newpage

\section{Introduction}

The flavor puzzle  of the Standard Model (SM) remains one of the 
outstanding issues in modern particle physics. The observed hierarchy
of the fermion masses is not explained in  the SM, but instead
parametrized in terms of small Yukawa couplings. One 
possibility that has been put forward independently in \cite{Babu:1999me} and \cite{Giudice:2008uua}
is that the Yukawa couplings are effective couplings dominated
by higher dimensional operators involving  the Higgs field. This 
eliminates small fundamental couplings in favor of ${\cal O}(1)$
parameters. The smallness of the fermion masses is then due to 
the smallness of the Higgs vacuum expectation value (VEV) compared
to the new physics scale  (of order TeV). This idea can be tested at the LHC 
by measuring the Higgs decay branching ratios.

In the present work, I study the conditions for CP violation to occur
in this framework. It is helpful to formulate the problem in a basis-independent
way, making use of  CP violating basis invariants which generalize
the Jarlskog invariant \cite{Jarlskog:1985ht,Bernabeu:1986fc,Gronau:1986xb}.
This approach has been employed in various new physics models, including 
the seesaw \cite{Branco:1986gr}, supersymmetric  \cite{Lebedev:2002wq}
and 2 Higgs doublet models \cite{Botella:1994cs}. 
The basis independent  formulation  allows one to obtain an invariant measure of CP violation,
which can be relevant to electroweak (EW)   baryogenesis 
(see \cite{Riotto:1999yt} for a review). Indeed, it is the smallness of the
Jarlskog invariant (and the Higgs sector constraints)  that makes  
electroweak baryogenesis essentially impossible in the SM.
The structure of the invariants with Higgs-dependent Yukawa couplings is very
different from that of the Standard Model and bears some similarity to the 
supersymmetric case \cite{Lebedev:2002wq}. It is therefore plausible that 
successful EW baryogenesis can be achieved.

\section{Higgs-dependent Yukawa couplings}

The main assumption of our framework  is that 
the Yukawa couplings are functions of the Higgs field.
They are effective quantities which  can be expanded in
powers of $ H^\dagger H /M^2 $ with $M$ being the cutoff
of the effective theory, 
\begin{equation}
Y_{ij}(H)= c^{(0)}_{ij}+ c^{(1)}_{ij} ~ {H^\dagger H \over M^2 }   + ...
\end{equation}
In the Standard Model, many of the Yukawa couplings are very small,
down to $10^{-5}$. It is therefore plausible that the higher
order terms are comparable or even dominant.
A particularly interesting possibility would be to have no small
fundamental couplings at all. For that the above expansion has to start
with some non-zero power of $ H^\dagger H /M^2 $. That is, all
$c_{ij}^{(n)}$ vanish until some integer $n_{ij}$,
\begin{equation}
Y_{ij}^{u,d}(H)= c_{ij}^{u,d} \left( {H^\dagger H \over M^2 } \right)^{n_{ij}^{u,d}} \;.
\label{yukawas}
\end{equation}
  This can happen due to symmetries
of the UV completion of our effective theory. 
Such symmetries are   particularly easy to realize  in supersymmetric (or 2 Higgs doublet model)
extensions of the Standard Model, where $ H^\dagger H $ is replaced by $H_1 H_2$.
The latter can have a (possibly discrete)   charge {\it \`a la } Froggatt-Nielsen 
\cite{Froggatt:1978nt}  such 
that the vanishing of some $c_{ij}^{(n)}$ is dictated by charge conservation.
In the SM case, the analog would be some non-abelian symmetry acting on $H^\dagger H$.\footnote{In general, one may separate the (low energy) 
Froggatt-Nielsen field and the Higgs (see, e.g. \cite{Tsumura:2009yf}).
In this case, the breaking of the Froggatt-Nielsen symmetry  {\it a priori}
has no relation to the electroweak scale.  }

In this framework, the smallness of the Yukawa couplings is explained by the smallness
of the Higgs VEV compared to the new physics scale $M$,
\begin{equation}
\epsilon \equiv {\langle H^\dagger H \rangle \over M^2} \simeq {1\over 60} \;,
\end{equation}
with the numerical value being fixed by $\epsilon = m_b/m_t$.
In terms of $\epsilon$, the Yukawa matrices are expressed as \cite{Giudice:2008uua}
\begin{equation}
Y^d \sim \left( \begin{matrix}
\epsilon^3 & \epsilon^2 & \epsilon^2 \\ 
\epsilon^3 & \epsilon^2 & \epsilon^2 \\ 
\epsilon^2 & \epsilon^1 & \epsilon^1
\end{matrix}   \right) ~~~,~~~
Y^u \sim \left( \begin{matrix}
\epsilon^3 & \epsilon^1 & \epsilon^1 \\ 
\epsilon^3 & \epsilon^1 & \epsilon^1 \\ 
\epsilon^2 & \epsilon^0 & \epsilon^0
\end{matrix}   \right)~~.   \label{texture}
\end{equation}
This texture reproduces the observed quark masses and mixings. 
The new feature is that the couplings to the physical Higgs boson
are modified dramatically. The relevant Lagrangian is given by
\begin{equation}
-{\cal L}= Y_{ij}^u (H)~ \bar q_{Li} u_{Rj} H^c + Y_{ij}^d (H)~ \bar q_{Li} d_{Rj} H  + {\rm h.c.}\;,
\end{equation}
where $H^c= i \sigma_2 H^*$ and  $Y_{ij}^{u,d} (H)$ are given by Eq.~(\ref{yukawas}).
The quark couplings to the physical Higgs increase by a  factor $2n_{ij}+1$ compared to that 
of the SM,
\begin{equation}
y_{ij}^{u,d} = (2n_{ij}^{u,d} +1) \bigl(  y_{ij}^{u,d}   \bigr)_{\rm SM} \;,
\end{equation} 
where $\bigl(  y_{ij}^{u,d}   \bigr)_{\rm SM}= m_{ij}^{u,d}/(\sqrt{2} v)  $ and 
the integers $n_{ij}^{u,d}$ can be read off from the texture (\ref{texture}).
As a result, the Higgs decay rate into quarks increases by a significant factor 
ranging from 9 for the bottom quark to 49 for up- and down- quarks leading to observable effects  at the LHC.

Since the mass matrices and the physical Higgs couplings differ by a flavor-dependent
factor, they cannot be diagonalized in the same basis and Higgs-mediated FCNC are induced.
These however are suppressed by the quark masses and, for the texture (\ref{texture}),
satisfy the experimental bounds (apart from $\epsilon_K$ which sets a mild constraint
on a CP phase). On the other hand, the flavor changing effects involving the top
quark are significant and can be observed at the LHC \cite{Giudice:2008uua}.

Along with extra flavor violation, this framework brings in additional sources of CP violation.
The extra CP phases reside in the quark couplings to the physical Higgs boson and can be 
relevant to baryogenesis. In what follows, I study the conditions for CP violation and 
construct the corresponding CP violating weak basis invariants.

\section{CP violating invariants with 2 quark species}

Consider a system of 2 quark species, say a top quark and a charm quark.
We have two relevant flavor objects: the mass matrix and the matrix
of the physical Higgs couplings. These are proportional to  
\begin{equation}
Y_{ij} ~~~~~,~~~~~ \tilde Y_{ij} \equiv  N_{ij}  Y_{ij} ~~,
\label{yukawa}
\end{equation}
respectively, with integer $ N_{ij}= 2  n_{ij} + 1 $.
This corresponds to  a special (``symmetric'') basis in which 
${\rm Arg} Y_{ij} ={\rm Arg} \tilde Y_{ij} $.
Under quark basis transformations $Y$ and $\tilde Y$ transform as
\begin{eqnarray}
&& Y \rightarrow U_L^\dagger ~Y~ U_R \;, \nonumber\\
&& \tilde Y \rightarrow U_L^\dagger ~\tilde Y~ U_R \;,
\label{transform}
\end{eqnarray}
where $U_L , U_R$ are unitary matrices. It is clear that, in general,
these matrices are not diagonal in the same basis. This provides us
with  sources for FCNC  and CP violation.  

It is easy to see that CP violation originates from a single CP phase.
Indeed, 3 complex phases in $Y$ and $\tilde Y$ can be eliminated
by a quark phase redefinition (\ref{transform}) with
\begin{eqnarray}
&& U_L = {\rm diag} (e^{i\alpha_1}, e^{i\alpha_2}) \;, \nonumber\\
&& U_R = {\rm diag} (e^{i\beta_1}, e^{i\beta_2}) \;. \label{phasefreedom}
\end{eqnarray}
In other words, CP violation is sourced by a reparametrization
invariant combination
\begin{equation}
 {\rm Im } \Bigl( ~Y_{11} Y_{22} Y_{12}^* Y_{21}^*~ \Bigr) \;. 
\end{equation}
This quantity  induces CP phases in the couplings of the quark
mass eigenstates to the physical Higgs. In the mass eigenstate
basis,
\begin{equation}  
Y \rightarrow \left( \begin{matrix}
y_1 & 0 \\ 
0 & y_2
\end{matrix}   \right) ~~~~,~~~~
\tilde Y \rightarrow \left( \begin{matrix}
\tilde{y}_{11} & \tilde{y}_{12} \\ 
\tilde{y}_{21} & \tilde{y}_{22}
\end{matrix}   \right) \;,
\label{mass-eigenstate}
\end{equation}
with positive $y_1$ and $y_2$ (proportional to the quark masses), 
the matrix of the Higgs couplings has 3 CP phases: 
Arg $\tilde y_{11}$, Arg $\tilde y_{22}$ and 
 Arg $\tilde y_{12} \tilde y_{21} $. Note that since this basis is only
defined up to a phase transformation $U_L=U_R= 
{\rm diag} (e^{i\delta_1}, e^{i\delta_2})$, the physical phases must be
invariant under this residual symmetry.

The presence of CP violation in the theory can be formulated in a basis
independent way.  For that one needs a quantity which is invariant under
the  U(2)$\times$U(2)  transformations  (\ref{transform}) and odd under the
CP transformation 
\begin{equation}
 Y \xrightarrow{\rm CP}  Y^* ~.
\end{equation}
Such invariants can be constructed systematically by forming an object
that transforms under one of the U(2)'s and taking a trace.\footnote{
This is analogous to  constructing gauge invariant operators \cite{Hanany:2010vu}. 
For instance, in the SM  
the flavor group   U(3)${}_L\times$U(3)${}_{R_u}\times$U(3)${}_{R_d}$ 
    can be gauged with 
$Y^u,  Y^d$ transforming as  bifundamentals. A choice of  $Y^u, Y^d$ breaks this 
symmetry {\it \`a la} Higgs, with 26 degrees of freedom being eaten by the 
SU(3)${}^3\times$U(1)${}^2$    gauge bosons (one U(1) is decoupled). 
The remaining 10 represent 
the observable masses, mixing angles and the CP phase. This also gives
 the dimension of the moduli space in the corresponding  SUSY gauge theory.}
For example, $Y \tilde Y^\dagger$ and $Y  Y^\dagger$ transform under
U${}_L$ only. Then, a trace of an anti-hermitian matrix formed out of these
objects will have the required properties. The simplest non-zero  
invariants are
\begin{eqnarray}
&& {\rm Tr} \bigl[ A^2 - {\rm h.c.}    \bigr] \;, \nonumber\\
&& {\rm Tr} \bigl[ AB - {\rm h.c.}    \bigr] \;,
\end{eqnarray}
where $A \equiv Y \tilde Y^\dagger$ and $B \equiv Y  Y^\dagger$.
Note that the invariant 
${\rm Tr} \bigl[ A - {\rm h.c.}    \bigr]$ vanishes identically for 
real $N_{ij}$. In terms of $Y_{ij}$ and $N_{ij}$, these invariants can
be expressed as
\begin{eqnarray}
&& {\rm Im~Tr} \bigl[ A^2  \bigr] = 2 (N_{12} N_{21} - N_{11} N_{22} )~
{\rm Im } \Bigl( ~Y_{11} Y_{22} Y_{12}^* Y_{21}^*~ \Bigr)   \;, \nonumber\\
&& {\rm Im ~Tr} \bigl[ AB     \bigr] = (N_{12} + N_{21} - N_{11}- N_{22})~
{\rm Im } \Bigl( ~Y_{11} Y_{22} Y_{12}^* Y_{21}^*~ \Bigr)  \;.
\end{eqnarray}
Note the appearance of the reparametrization invariant quantity
${\rm Im } \bigl( ~Y_{11} Y_{22} Y_{12}^* Y_{21}^*~ \bigr) $. 
If it is zero, all CP odd invariants vanish. 
Since there is only one independent CP phase, the vanishing of one 
CP odd invariant  in the non-degenerate case 
guarantees that there is no CP violation.

In the degenerate case, i.e. when there are special relations among 
$N_{ij}$'s or eigenvalues,  the situation is more subtle.
For example, ${\rm Im~Tr} \bigl[ A^2  \bigr]$ vanishes if 
Det$~N=0$. Yet, ${\rm Im~Tr} \bigl[ AB  \bigr]$ can be non-zero.
However, if both vanish, no CP violation is possible.
To see this, note that Det$~N=0$ means  that the columns (or rows)
of $N$ are linearly dependent, which  in conjunction with
 $N_{12} + N_{21} - N_{11}- N_{22}=0$ implies that $N$
has the form
\begin{equation}
N  =  \left( \begin{matrix}
N_1 & N_2 \\ 
N_1 & N_2
\end{matrix}   \right) \;, \label{degen}
\end{equation}
up to a transposition. Then, $\tilde Y$ factorizes as
\begin{equation} 
\tilde Y = Y~ {\rm diag} (N_1, N_2) \;.
\end{equation}
In the mass eigenstate basis (\ref{mass-eigenstate}), it has 
the form
\begin{equation}
\tilde Y ~\rightarrow~ {\rm diag} (y_1, y_2)~ 
U_R^{\dagger} \;{\rm diag} (N_1, N_2) \; U_R
\;.
\end{equation}
Since $U_R^{\dagger} {\rm diag} (N_1, N_2)  U_R$ is hermitian and $y_{1,2}$
real, the only phase of the resulting matrix can be removed by 
phase redefinition with $U_L=U_R= 
{\rm diag} (e^{i\delta_1}, e^{i\delta_2})$. 
In other words,
Arg $\tilde y_{11}$, Arg $\tilde y_{22}$ and  
 Arg $\tilde y_{12} \tilde y_{21} $ all vanish. 
Since the flavor objects are real in this basis, all possible CP violating
invariants vanish. 

The reparametrization  invariant
${\rm Im } \bigl( ~Y_{11} Y_{22} Y_{12}^* Y_{21}^*~ \bigr) $
can vanish due to hidden symmetries. In the mass eigenstate basis,
it can be written as
\begin{equation}
{\rm Im } \bigl( ~Y_{11} Y_{22} Y_{12}^* Y_{21}^*~ \bigr)=
y_1 y_2 (y_1^2-y_2^2) ~ {\rm Im} \bigl(  U_{L_{11}}^*   U_{L_{12}}
U_{R_{11}}
   U_{R_{12}}^*       \bigr) \;,
\end{equation}
where $y_{1,2}$ are the eigenvalues of $Y$ and $U_L^\dagger Y U_R = {\rm
diag}(y_1,y_2) $.
It is then clear that there can be no CP violation if there is a massless
eigenstate
or degenerate spectrum. In the latter case, there is an extra U(2) symmetry
which
eliminates the CP phase. Similarly, there is an extra U(1) associated with
phase
redefinition of the massless state. This is qualitatively different from CP violation in the Standard
Model.
Recall that only in the degenerate (and not in the massless)
case can one rotate away the CKM phase. This has to do with the fact that CP
violation
in the SM is associated with the relative phases in $Y^u Y^{u \dagger }$ and
     $Y^d Y^{d \dagger }$ which both transform under $U_L$, whereas in our
case
CP violation is due to the phases between $Y$ and $\tilde Y$ which transform
under biunitary transformations $U_L$ and $U_R$.

\section{Generalizations}

\subsection{3 flavor case}

Although CP violation comes predominantly from the mixing of 2 flavor states, 
it is instructive to consider the 3 flavor case.
The Yukawa matrix has 9 phases, 5 of which can be eliminated by quark
phase redefinitions leaving 4 physical phases. These can be chosen as
\begin{equation} 
{\rm Arg} \bigl(~    Y_{ij} Y_{i+1,j+1} Y_{i+1,j}^* Y_{i,j+1}^*  ~  \bigr)
\label{3X3phases}
\end{equation}
with $i,j=1,2$.
The corresponding weak basis invariants can be taken to be
\begin{eqnarray}
&& {\rm Tr} \bigl[ A^k - {\rm h.c.}    \bigr] \;, \nonumber\\
&& {\rm Tr} \bigl[ A^l B^m - {\rm h.c.}    \bigr] \;,  \label{3X3invariants}
\end{eqnarray}
with $A \equiv Y \tilde Y^\dagger$ and $B \equiv Y  Y^\dagger$ and
integer $k,l,m$ ($k>1$).
In the non-degenerate case, the vanishing of 4 independent invariants would
ensure absence of CP violation.\footnote{The resulting equations are non-linear
in CP phases and may have spurious solutions for special values of 
the mixing angles. Here we ignore this possibility 
(for a related discussion, see \cite{Esmaili:2007av}). }

The degenerate case is rather complicated. For a special class 
of $N$-matrices, no CP violation is possible.
The analog of Eq.~(\ref{degen}) is
\begin{equation}
N  =  \left( \begin{matrix}
N_1 & N_2 & N_2\\ 
N_1 & N_2 & N_2\\
N_1 & N_2 & N_2
\end{matrix}   \right) \;, 
\end{equation}
up to a transposition and permutations of the columns.
In this case, $\tilde Y$ has the following form in the mass eigenstate basis:
\begin{equation}
\tilde Y ~\rightarrow~ {\rm diag} (y_1, y_2, y_3)~ 
U_R^{\dagger} \;{\rm diag} (N_1, N_2, N_2) \; U_R
\;.
\end{equation}
The only reparametrization invariant phase of the hermitian matrix 
 $U_R^{\dagger} {\rm diag} (N_1, N_2, N_2) U_R$ can be removed due to
the U(2) symmetry of the lower $2\times 2$ block. 
Note that it is not sufficient to have a rank 1 structure and 2 columns
of $N$ must be identical to ensure absence of CP violation.
Unlike in the  $2\times 2$ case, 
it is not clear what is the minimal set of CP odd invariants, vanishing of
which would ensure  absence of CP violation since the resulting equations are 
highly non-linear in $N_{ij}$.

\subsection{Inclusion of up- and down-sectors} 

\subsubsection{2 generations}

A different class of  CP violating phases result from an interplay
of the up- and down- sectors with the symmetry group 
U(2)${}_L\times$U(2)${}_{R_u}\times$U(2)${}_{R_d}$.
 In our framework, we have 
4 flavor objects $Y^u, \tilde Y^u, Y^d, \tilde Y^d$ with the following
transformation properties :
\begin{eqnarray}  
&& Y^u \rightarrow U_L^\dagger ~Y^u~ U_{R_u} ~~,~~ 
   \tilde Y^u \rightarrow U_L^\dagger ~\tilde Y^u~ U_{R_u} \;, 
\nonumber\\
&& Y^d \rightarrow U_L^\dagger ~Y^d~ U_{R_d} ~~,~~
   \tilde Y^d \rightarrow U_L^\dagger ~\tilde Y^d~ U_{R_d} \;,
\end{eqnarray}
as required by the SU(2)${}_L$ symmetry.
Out of these matrices one can form various objects that transform
under  one of the symmetries. For instance, 
$Y^{u\dagger} Y^u $, $Y^{u\dagger} \tilde Y^u $ and  
$\tilde Y^{u\dagger} \tilde Y^u $ all transform under  $U_{R_u}$.
Their misalignment results in the CP phase studied in the previous 
section. On the other hand, quantities transforming under  $U_L$
involve both up- and down- sectors: 
$Y^u Y^{u\dagger} $, $Y^{d} Y^{d\dagger} $, etc.   
They are responsible for the extra CP phases.

Consider our ``symmetric'' basis  (\ref{yukawa}). 
5 out of 8 phases in  $Y^u, Y^d$ can be eliminated by
\begin{eqnarray}
&& U_L = {\rm diag} (e^{i\alpha_1}, e^{i\alpha_2}) \;, ~
 U_{R_u} = {\rm diag} (e^{i\beta_{1u}}, e^{i\beta_{2u}}) \;,~
U_{R_d} = {\rm diag} (e^{i\beta_{1d}}, e^{i\beta_{2d}}) \;.
\end{eqnarray}
The 3 physical phases can be chosen as
\begin{eqnarray}
  \phi_u &=& {\rm Arg } \Bigl( ~Y_{11}^u Y_{22}^u Y_{12}^{u*} Y_{21}^{u*}~ \Bigr) \;, \nonumber \\
  \phi_d &=& {\rm Arg } \Bigl( ~Y_{11}^d Y_{22}^d Y_{12}^{d*} Y_{21}^{d*}~ \Bigr) \;, \nonumber \\
 \phi &=& {\rm Arg } \Bigl( ~Y_{11}^u Y_{21}^{u*} Y_{11}^{d*} Y_{21}^{d}~ \Bigr) \;. \label{fi}
\end{eqnarray}
The phase $\phi$ is a new object resulting from a misalignment of the 
two sectors.
The corresponding  CP violating basis invariants 
(in a non-degenerate case) are
\begin{eqnarray}
&& {\rm Tr} \bigl[ (Y^u \tilde Y^{u\dagger} )^2 - {\rm h.c.}  \bigr] 
\;,\nonumber \\
&& {\rm Tr} \bigl[ (Y^d \tilde Y^{d\dagger} )^2 - {\rm h.c.}  \bigr] 
\;,\nonumber \\
&& {\rm Tr} \bigl[Y^u Y^{u\dagger},  \tilde Y^u \tilde Y^{u\dagger}   , Y^d Y^{d\dagger}    \bigr]     \;,        \label{3invariants}
\end{eqnarray}
where $[A,B,C]$ denotes a completely antisymmetric product of $A,B$ and $C$.
While the first two invariants are proportional to  $\sin \phi_u$ and $\sin\phi_d$, the last one is sensitive to $\sin\phi$. Invariants of this type have 
appeared before in the context of supersymmetry \cite{Lebedev:2002wq}. 
Note that
there is no Jarlskog-type invariant since 
${\rm Tr} [Y^u Y^{u\dagger}, Y^d Y^{d \dagger}]^3=0$ for 
2 generations.

An explicit calculation gives
\begin{eqnarray}
 {\rm Tr} \bigl[Y^u Y^{u\dagger},  \tilde Y^u \tilde Y^{u\dagger}   , Y^d Y^{d\dagger}    \bigr] &=& i a \sin \phi_u + i b \sin\phi + i c \sin(\phi+\phi_d)
\nonumber\\
&+& i d \sin(\phi-\phi_u) +ie \sin(\phi +\phi_d - \phi_u) \;, 
\end{eqnarray}
where 
\begin{eqnarray}
&& a= 6 f(Y^d) 
~(N_{12} N_{22} -N_{11}N_{21})
~\big\vert Y_{11}^u Y_{12}^u Y_{21}^u Y_{22}^u \big\vert\;, \nonumber\\
&&  b=6 \Bigl(  f(Y^u) N_{11}N_{21}- f(\tilde Y^u)  
\Bigr) ~\big\vert Y_{11}^u Y_{21}^u Y_{11}^d Y_{21}^d \big\vert \;, \nonumber\\
&&  c=6 \Bigl(  f(Y^u) N_{11}N_{21}- f(\tilde Y^u)  
\Bigr) ~\big\vert Y_{11}^u Y_{21}^u Y_{12}^d Y_{22}^d \big\vert \;, \nonumber\\
&&  d=6 \Bigl(  f(Y^u) N_{12}N_{22}- f(\tilde Y^u)  
\Bigr) ~\big\vert Y_{12}^u Y_{22}^u Y_{11}^d Y_{21}^d \big\vert \;, \nonumber\\
&&  e=6 \Bigl(  f(Y^u) N_{12}N_{22}- f(\tilde Y^u)  
\Bigr) ~\big\vert Y_{12}^u Y_{22}^u Y_{12}^d Y_{22}^d \big\vert \;, \nonumber
\end{eqnarray}
and $f(Y)$ is defined by
\begin{equation}
 f(Y) \equiv  \vert Y_{11} \vert^2 +\vert Y_{12} \vert^2 -
\vert Y_{21} \vert^2  - \vert Y_{22} \vert^2 \;.
\end{equation}
We see that this invariant is controlled by  $\sin\phi$.
In the non-degenerate case, the vanishing of the 3 invariants 
(\ref{3invariants})  implies 
absence of CP violation. The first 2 invariants are proportional to 
$\sin \phi_u$ and $\sin \phi_d$, respectively, while for $\phi_u=\phi_d=0$,
the last invariant is proportional to $\sin\phi$.  

CP violation in this system exists even if the $N$-matrix has the 
degenerate form (\ref{degen}) in both sectors, as long $N_1 \not= N_2$. 
In this case, 
${\rm Tr} \bigl[Y^u Y^{u\dagger},  \tilde Y^u \tilde Y^{u\dagger}   , Y^d Y^{d\dagger}    \bigr]$ generally does not vanish. There is no CP violation
if $N$ has equal matrix elements, as it should be, since all the 
coefficients $a$ to $e$ vanish. 

It is instructive to consider the above invariant in the mass eigenstate 
basis.  Diagonalizing $Y^u Y^{u\dagger} \rightarrow {\rm diag}(y_1^{u2},
y_2^{u2})$ and parametrizing 
$\tilde Y^u \tilde Y^{u\dagger}= U {\rm diag}(\tilde y_1^{u2}, \tilde
y_2^{u2}) U^\dagger$, 
$Y^d Y^{d\dagger} = V {\rm diag}(y_1^{d2},
y_2^{d2}) V^\dagger$ in this basis, we have
\begin{equation}
{\rm Tr} \bigl[Y^u Y^{u\dagger},  \tilde Y^u \tilde Y^{u\dagger}   , Y^d Y^{d\dagger}    \bigr]= 6i (y_1^{u2}-y_2^{u2}) (\tilde y_1^{u2}- \tilde y_2^{u2})
(y_1^{d2}-y_2^{d2}) {\rm Im} \bigl(  U_{11} U_{21}^* V_{11}^* V_{21}       \bigr) \;. \label{K-inv}
\end{equation}
More generally, for hermitian $A,B$ and $C$, the invariant  Tr$[A,B,C]$ is proportional to the sine of 
Arg$(B_{12} C_{12}^*)$, which  is the only  reparametrization invariant phase in this basis 
(the basis is defined up to $U_L={\rm diag}(e^{i\delta_1},e^{i\delta_2})  $).
If the invariant vanishes, the phase is zero (or can be rotated away)  
and no other CP violating invariant  out of $A,B$ and $C$ can be constructed.

Consider the degenerate case. The above invariant vanishes if there are degenerate eigenvalues.
In this case, the residual symmetry  is U(2) instead of U(1) and two matrices  can be diagonalized
simultaneously.
Therefore, all objects can be made real in this basis   and all CP violating invariants vanish.
Another possible degeneracy lies in the  $N$-matrix. If all matrix elements of $N$ are the same,
two matrices can be diagonalized simultaneously and no CP violation occurs. This is not generally the 
case for $N$-matrices of the form  (\ref{degen}) with $N_1 \not= N_2$, and there is CP violation. 

We have so far considered the system of $Y^u Y^{u\dagger},  \tilde Y^u \tilde Y^{u\dagger}$ and $ Y^d Y^{d\dagger} $. 
The discussion can be repeated for other choices of the 3 hermitian objects, including
$ \tilde Y^d \tilde Y^{d\dagger} $. It is easy to see that if both $Y^u$ (or $\tilde Y^u$)  and 
$Y^d$ (or $\tilde Y^d$) 
have degenerate eigenvalues, no CP violation is possible.
This is also the case when $N$ in the up sector has identical matrix elements and, at the same time,
  $N$ in the down  sector has the same property.

\subsubsection{3 generations}

The generalization to the case of 3 generations is straightforward.
The symmetry group is U(3)${}_L\times$U(3)${}_{R_u}\times$U(3)${}_{R_d}$.
Out of 18 phases of $Y_{u,d}$,  3$\times$3-1=8 can be eliminated, leaving 10 physical. 4+4=8 of them have the form
(\ref{3X3phases}), while the last two are analogs of $\phi$ in (\ref{fi}),
\begin{equation}
{\rm Arg } \Bigl( ~Y_{11}^u Y_{21}^{u*} Y_{11}^{d*} Y_{21}^{d}~ \Bigr) ~~,~~
{\rm Arg } \Bigl( ~Y_{22}^u Y_{32}^{u*} Y_{22}^{d*} Y_{32}^{d}~ \Bigr)~.
\end{equation}
In addition to the invariants  (\ref{3X3invariants}) in each sector,
one can take further two  of the form 
\begin{equation}
{\rm Tr } \bigl[ A^k,B^l,C^m           \bigr]  \label{3newinv}
\end{equation} 
as the invariants sensitive to the above 2 phases. Here $k,l,m$ are integer and $A,B,C$ are 
hermitian matrices from the list \{$Y^u Y^{u\dagger}, \tilde Y^u \tilde Y^{u\dagger},
Y^d Y^{d\dagger}, \tilde Y^d \tilde Y^{d\dagger} $\}.\footnote{One can form further
hermitian matrices, but this list would suffice in the non-degenerate case. }
This completes the list of 10 relevant CP violating invariants  
in the non-degenerate case.

Some discussion of the degenerate case can be found in Ref.\cite{Lebedev:2002wq}.  
Clearly, CP violation exists even if the $N$-matrices have identical  elements.
In this case there is a CKM phase and the Jarlskog invariant replaces 
(\ref{3newinv}). Also, there have to be many degenerate  eigenvalues to eliminate 
CP violation. For example, even if $Y^u$ and $Y^d$ have all degenerate eigenvalues,
one can still form the Jarlskog-type invariant
${\rm Tr} [\tilde Y^u \tilde Y^{u\dagger}, \tilde Y^d \tilde Y^{d \dagger}]^3$,
which would  not vanish in general. A complete study of the degenerate case is beyond
the scope of this work.

\section{Applications}

In the Standard Model, CP violation is controlled by the Jarlskog invariant
of order 12 in quark masses,
\begin{eqnarray}
{\rm Im ~Tr} [Y^u Y^{u\dagger}, Y^d Y^{d \dagger}]^3 &=& (y_u^2 -y_c^2) (y_c^2 -y_t^2)
 (y_t^2 -y_u^2) (y_d^2 -y_s^2) (y_s^2 -y_b^2) (y_b^2 -y_d^2) \nonumber\\
&\times &  J ~ \sim~ 10^{-22}\;, \label{jarlskog}
\end{eqnarray}
where $J \sim 10^{-5}$ is a combination of the CKM matrix entries.  
Its smallness is due to the fact that one must have at least  3 generations and both
up- and down-sectors in order to have CP violation. 
In the context of electroweak baryogenesis, the Jarlskog invariant   appears in the calculation
 of the CP asymmetry \cite{Shaposhnikov:1986jp}, with the factor $v^{12}/T^{12} \sim 1$, where $v=174$ GeV and $T$ is
the temperature of the electroweak transition.  
One of the problems with the SM baryogenesis is that it is very difficult if not impossible
to generate the observed baryon asymmetry $\eta \sim 10^{-10}$ out of such a small number (see 
however \cite{Tranberg:2009de}).

If one allows for Higgs--dependent Yukawa couplings, 
the situation changes dramatically. CP violation exists already for 2 generations within 
a single  (up- or down-) sector. 
The corresponding weak basis invariant is of order 4 in quark masses. Taking
as an example a system of a top and an up quark, the invariant is given by
\begin{eqnarray}
{\rm Im ~Tr} \bigl[ \bigl( Y^u \tilde Y^{u\dagger} \bigr)^2 \bigr] &=& y_t y_u (y_u^2 - y_t^2) 
\nonumber\\
&\times& 2~ {\rm Det } N~ {\rm Im} \bigl(  U_{L_{11}}^*   U_{L_{12}}
U_{R_{11}}
   U_{R_{12}}^*       \bigr) \sim 10^{-9}~ \sin\delta \;,   \label{numer}
\end{eqnarray}
where $\delta$ is the relevant CP phase. It is more than 10 orders of magnitude larger
than the Jarlskog invariant which is likely to be  sufficient to generate the required baryon
asymmetry. 

Let us elaborate on the above calculation. To obtain this number, it is necessary to
assume a specific Yukawa texture. 
Inspection of the texture (\ref{texture}) shows that there is no CP violation in the \{$t,c$\} 
system because it falls into the ``degenerate'' category (\ref{degen}). However,
in the \{$t,u$\} system CP violation exists. The relevant 2$\times$2  Yukawa texture and 
the $N$-matrix are
\begin{equation}
Y^u \sim \left( \begin{matrix}
\epsilon^3 & \epsilon^1 \\ 
\epsilon^2 & \epsilon^0
\end{matrix}   \right) ~~~,~~~
N=\left( \begin{matrix}
7 & 3 \\ 
5 & 1
\end{matrix}   \right) ~~~.
\end{equation}
In the mass eigenstate basis, the matrix of the Higgs couplings $\tilde Y^u$ is not
diagonal and contains complex phases.  
The diagonalizing matrices have the form
\begin{equation}
V_L \sim \left( \begin{matrix}
1 & -\epsilon \\ 
\epsilon & 1
\end{matrix}   \right) ~~~,~~~
V_R \sim \left( \begin{matrix}
1 & -\epsilon^2 \\ 
\epsilon^2 & 1
\end{matrix}   \right)~~,
\end{equation}
such that 
\begin{equation}
\tilde Y^u \sim \left( \begin{matrix}
\epsilon^3 & \epsilon^1 \\ 
\epsilon^2 & \epsilon^0
\end{matrix}   \right) \label{ytilde}
\end{equation}
in the mass eigenstate basis. Its off-diagonal elements carry order 
one phases. 
The combination ${\rm Im} \bigl(  U_{L_{11}}^*   U_{L_{12}}
U_{R_{11}}   U_{R_{12}}^*       \bigr) $ is of order $\epsilon^3$,
which gives  the estimate  (\ref{numer}). 
Of course, the main difference between Eqs.~(\ref{jarlskog}) and (\ref{numer})
is the absence of the large quark mass suppression in the latter,
 which is independent of the Yukawa texture.

\begin{figure}
\hspace*{4cm}
\includegraphics[width=6.0cm]{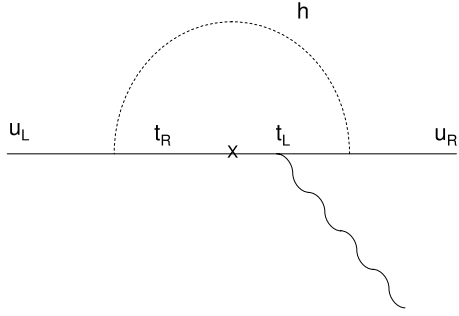}
\vspace*{0.3cm}
 \caption{\footnotesize The leading contribution to the neutron EDM.  }
\label{f1} 
\end{figure}

The increase in the amount of CP violation does not come for free.
The same effect generates the neutron EDM at one loop. This places a 
 constraint on the relevant CP phase. The leading contribution comes
from the flavor off-diagonal $t-u$ interactions (Fig.~\ref{f1}), 
which induce the 
neutron EDM at order $\epsilon^3$. A simple estimate shows that the 
reparametrization invariant  CP phase Arg$\bigl( \tilde Y^u_{12} \tilde Y^u_{21} \bigr)$ 
has to be smaller than $10^{-1}$. (Of course, the estimate depends 
on the ``order 1'' coefficients in the texture and  reducing the off-diagonal
entries helps relax the bound). As a result, the phase $\delta$ in 
Eq.(\ref{numer}) cannot be greater than $10^{-1}$.  Nevertheless, 
the value of the invariant is still sufficiently large to be compatible
with the observed baryon asymmetry. A similar in spirit study of EDMs versus 
EW baryogenesis can be found in \cite{Huber:2006ri}.

One can also use other CP-violating invariants involving both up- and down-
 sectors.
Consider 2 heavy generations.  According to 
Eq.(\ref{K-inv}),
\begin{equation}
{\rm Im~ Tr} \bigl[Y^u Y^{u\dagger},  \tilde Y^u \tilde Y^{u\dagger}   , Y^d Y^{d\dagger}    \bigr] \sim y_t^4 y_b^2 ~ U_{21} V_{21} ~\sin\phi \;.
\end{equation}
For our texture (\ref{texture}), this is of order $\epsilon^4 \sin\phi \sim 10^{-7} 
\sin\phi$. 
Note that  CP violation  exists in this system  despite the degenerate $N$-matrix for the  $t-c$ block.
The corresponding CP phase $\phi$ is essentially unconstrained
because the FCNC bounds from  the heavy quark  systems are satisfied for any
phase, while the EDM contribution comes at two loops. 
It is interesting that 
 rephasing invariance requires  interference with
the SM contribution mediated by the $W$ boson. In the mass eigenstate 
basis  $Y^u= {\rm diag} (y_t, y_c)$, $Y^d= {\rm diag} (y_b, y_s)$,  
the CKM phase convention eliminates the residual phase symmetry
$U_L^u=U_R^u$ and $U_L^d=U_R^d$.
For example, consider the $t-c$ flavor change.
 While the Higgs exchange generates 
operators like $(\bar t_L c_{R})^2  $, the $W$ exchange
generates $(\bar t_L  c_L)^2 $. The physical phase 
between them is fixed by requiring real masses and real $W$-vertices.
 For the light generations,
an analogous physical phase is constrained by $\epsilon_K$  \cite{Giudice:2008uua}.

An insufficient amount of CP violation is not the only obstacle for 
baryogenesis in the SM. The other problem is that it fails to provide  
a sufficiently strong first order phase transition, which would only be  
possible  for an unacceptably light Higgs,    $m_h < 72$ GeV.
As a result, the baryon asymmetry
is erased by the sphaleron processes.  
This statement is no longer true if there is a dimension six operator
\cite{Zhang:1992fs,Grojean:2004xa,Bodeker:2004ws}
\begin{equation}
\Delta V = {1\over \Lambda^2}~ \bigl(  H^\dagger H - v^2      \bigr)^3,
\end{equation}
with $\Lambda \sim 1$ TeV. This operator changes the relation between the
strength of the EW phase transition and the Higgs mass such that
EW baryogenesis  becomes possible.
In our framework, such an operator is expected to be generated by integrating
out TeV mass  states, for instance, gauge singlets. The details depend
on a particular UV completion of our effective theory, but the presence 
of the above operator is ``decoupled'' from the CP and flavor physics,
and can safely be assumed.

To summarize, it appears that the framework of Higgs-dependent Yukawa couplings has
the necessary ingredients to address the problem of baryogenesis. 
In particular, the Higgs interactions with the top and up quarks contain
a sufficient amount of CP violation.
A detailed  study will be presented elsewhere.

\section{Conclusion}

The framework of Higgs-dependent Yukawa couplings allows one to eliminate
small fundamental couplings from the Standard Model. In this work, I have
analyzed  the conditions for CP violation to occur in such a setup. In particular,
I have identified a class of  basis invariants responsible for CP violation.
Unlike in the Standard Model, the CP symmetry can already be violated in a system
of 2 quark species.
The invariant measure of CP violation is found to be   more than 10 orders of magnitude
greater than that in the Standard Model. It is therefore plausible that 
 this framework contains
a sufficient amount of CP violation for successful electroweak baryogenesis.

{\bf Acknowledgements.} I am grateful to L. Lebedeva for enlightening discussions and
motivation.

\end{document}